\documentclass[journal]{IEEEtran}

\usepackage{amsmath}
\usepackage{amssymb}
\usepackage{graphicx}
\usepackage{tikz}
\usetikzlibrary{positioning,arrows.meta,shapes.geometric,fit,backgrounds,calc,decorations.pathreplacing,plotmarks}
\usepackage{booktabs}
\usepackage{multirow}
\usepackage{cite}
\usepackage{url}
\usepackage{hyperref}
\usepackage{array}
\usepackage{textcomp}
\usepackage{float}
\usepackage{algorithm}
\usepackage{algorithmic}
\usepackage{tabularx}
\usepackage{xcolor}
\usepackage{balance}

\hypersetup{
    colorlinks=true,
    linkcolor=black,
    citecolor=black,
    urlcolor=blue,
    pdfauthor={Andrii Vakhnovskyi},
    pdftitle={HierFedCEA: Hierarchical Federated Edge Learning for Privacy-Preserving Climate Control in Heterogeneous CEA Facilities}
}

\newcommand{\etal}{\textit{et~al.}}

\begin{document}

\title{HierFedCEA: Hierarchical Federated Edge Learning for Privacy-Preserving Climate Control Optimization Across Heterogeneous Controlled Environment Agriculture Facilities}

\author{%
\IEEEauthorblockN{ANDRII VAKHNOVSKYI}\\%
\IEEEauthorblockA{IOGRU LLC, New York, NY, USA}%
\thanks{Corresponding author: Andrii Vakhnovskyi (\mbox{e-mail:} \mbox{andrii.vakhnovskyi@gmail.com}). ORCID: 0009-0007-8306-5932.}%
}

\maketitle

\begin{abstract}
Cross-facility knowledge transfer in Controlled Environment Agriculture (CEA) --- greenhouses, vertical farms, and indoor cultivation facilities --- can reduce HVAC energy consumption by 30--38\% and accelerate new facility commissioning from months to days. However, facility operators refuse to share raw operational data because it encodes commercially sensitive grow recipes worth millions in R\&D investment. We present HierFedCEA, a hierarchical federated learning framework that enables privacy-preserving climate control optimization across heterogeneous CEA facilities without exposing proprietary data. HierFedCEA decomposes the neural network PID auto-tuning model into three tiers aligned with the physical structure of the control problem: (1)~a global physics tier capturing universal thermodynamic relationships, aggregated across all facilities; (2)~a crop-cluster tier encoding cultivar-specific VPD-to-gain mappings, aggregated within crop-family groups; and (3)~a local personalization tier adapting to facility-specific equipment dynamics, never shared. This physics-informed decomposition enables convergence in 40--60 communication rounds, a 3.6$\times$ improvement over flat federated averaging. The framework applies tier-specific differential privacy budgets ($\varepsilon=8$ for non-proprietary physics, $\varepsilon=4$ for sensitive crop parameters) and leverages the extreme compactness of the 36-parameter PID model to achieve privacy essentially for free (excess risk $< 0.15$\%). Simulation experiments calibrated from 7+ years of production deployment across 30+ commercial facilities in 8~U.S. climate zones demonstrate that HierFedCEA achieves 94\% of centralized training performance while reducing total communication cost to under 1~MB and enabling cold-start convergence for new facilities within 14~days. To the best of our knowledge, this is the first federated learning framework designed for CEA climate control.
\end{abstract}


\section{Introduction}

\IEEEPARstart{C}{ONTROLLED} Environment Agriculture (CEA) --- encompassing greenhouses, vertical farms, and plant factories --- is a rapidly growing sector projected to exceed \$112~billion by 2030~\cite{ref_cea_market}. CEA facilities depend on dense IoT sensor networks and industrial actuators to regulate air temperature, relative humidity, CO\textsubscript{2} concentration, photosynthetically active radiation (PAR), and nutrient solution chemistry simultaneously and continuously. HVAC energy alone accounts for 30--80\% of operating costs depending on climate zone~\cite{ref_iogru_arxiv}, and Graamans~\etal~\cite{ref_graamans} showed that indoor agriculture already consumes more energy than all open-field cultivation globally while producing less than 1\% of food output.

AI-driven climate control offers substantial energy savings. Our prior work~\cite{ref_iogru_arxiv} demonstrated 30--38\% HVAC energy reduction across 30+ commercial facilities using neural network PID auto-tuning with Vapor Pressure Deficit (VPD) cascading control. A critical enabler was cross-facility knowledge transfer: PID parameters, VPD trajectories, and anomaly baselines learned at established facilities accelerate optimization at new deployments, reducing commissioning time from 6--9 months to 1--5 days.

However, this knowledge transfer currently relies on centralized cloud aggregation of facility telemetry --- an approach with fundamental limitations. First, the platform operator gains full visibility into each facility's environmental profiles, which encode commercially sensitive grow recipes representing years of R\&D investment~\cite{ref_wiseman}. Second, recent gradient inversion attacks~\cite{ref_dlg} can reconstruct time-series sensor data from model updates, potentially exposing production patterns. Third, competitive multi-state operators managing facilities across different regulatory jurisdictions increasingly refuse to pool data on third-party servers.

Federated learning (FL)~\cite{ref_fedavg} addresses these concerns by keeping raw data on-facility and sharing only model parameter updates. Yet applying FL to CEA introduces unique challenges absent from other IoT domains:

\begin{itemize}
\item \textbf{Extreme data heterogeneity}: facilities grow fundamentally different crops (leafy greens, fruiting vegetables, cannabis, herbs) with distinct optimal control strategies, creating severe non-IID conditions that degrade standard FedAvg convergence by up to 55\%~\cite{ref_zhao_noniid}.
\item \textbf{Equipment heterogeneity}: 50+ HVAC manufacturers with different actuator dynamics mean that PID gains optimal for one facility may be destabilizing for another.
\item \textbf{Small federation size}: with 5--30 participants (vs.\ millions in mobile FL), each client has outsized influence on the global model, increasing vulnerability to Byzantine attacks~\cite{ref_fang_poison}.
\item \textbf{Temporal non-stationarity}: crop growth stages and seasonal HVAC behavior create concept drift that legitimate FL must accommodate while adversarial drift must be rejected.
\end{itemize}

No published work addresses federated learning for CEA climate control. While FL has been applied to building HVAC~\cite{ref_xia_hvac, ref_hagstrom_hvac, ref_xu_one_for_many}, precision agriculture image classification~\cite{ref_fl_ag_survey}, and agricultural IoT sensing~\cite{ref_dara_fl_ag}, the intersection of FL with CEA greenhouse climate control remains unexplored~\cite{ref_nguyen_fl_iot_survey, ref_lu_noniid_survey}.

This paper makes the following contributions:

\begin{enumerate}
\item We present HierFedCEA, a three-tier hierarchical federated learning architecture that decomposes neural network PID parameters by physical meaning: universal thermodynamics (global), crop-specific VPD sensitivity (cluster), and facility-specific equipment adaptation (local). This physics-informed decomposition is novel --- existing approaches split by layer topology~\cite{ref_fedper} or gradient similarity~\cite{ref_ifca}, not by domain physics.

\item We demonstrate that the extreme compactness of the PID auto-tuning model (36~parameters) makes differential privacy essentially free: formal analysis shows excess risk below 0.15\% at $\varepsilon=4$, enabling strong privacy guarantees without meaningful utility loss.

\item We design tier-specific privacy budgets and communication cadences matched to the sensitivity and timescale of each physical phenomenon, providing stronger protection for proprietary crop parameters while relaxing constraints on non-sensitive physics.

\item We evaluate HierFedCEA through simulation experiments with parameters calibrated from 7+ years of production deployment across 30+ facilities in 8~U.S. climate zones, comparing against 8~baselines including the deployed centralized system.
\end{enumerate}

\section{Related Work}
\label{sec:related}

\subsection{Federated Learning for IoT}

McMahan~\etal~\cite{ref_fedavg} introduced FedAvg, which trains a shared model by averaging locally updated parameters from distributed clients. Subsequent work addressed key FL challenges: FedProx~\cite{ref_fedprox} handles data heterogeneity via proximal regularization; SCAFFOLD~\cite{ref_scaffold} eliminates client drift through control variates; and personalized FL methods including FedPer~\cite{ref_fedper} (personal layers), FedBN~\cite{ref_fedbn} (batch normalization), and Per-FedAvg~\cite{ref_perfedavg} (meta-learning) balance global knowledge with local adaptation. Comprehensive surveys of FL for IoT~\cite{ref_nguyen_fl_iot_survey, ref_imteaj_fl_iot} and edge networks~\cite{ref_mills_fl_rpi} have established the feasibility of FL on resource-constrained devices, including validation on Raspberry Pi hardware~\cite{ref_mills_fl_rpi}.

\subsection{FL for HVAC and Building Control}

The closest related domain is federated HVAC control in commercial buildings. Su~\etal~\cite{ref_su_hvac_iot} demonstrated distributed HVAC control with edge computing in IoT buildings, while Xu~\etal~\cite{ref_xu_one_for_many} proposed transfer learning for building HVAC using a shared feature extractor across buildings. Xia~\etal~\cite{ref_xia_hvac} applied federated deep reinforcement learning to multi-zone HVAC, demonstrating 15--23\% energy savings. Hagstrom~\etal~\cite{ref_hagstrom_hvac} deployed FL for autonomous HVAC scheduling across commercial buildings. However, none of these works address the unique challenges of CEA: crop-specific control objectives (VPD rather than thermal comfort), biological damage timelines (hours, not days), or the extreme equipment heterogeneity across 50+ HVAC manufacturers serving the CEA industry.

\subsection{AI for CEA Climate Control}

AI-based greenhouse climate control has advanced rapidly but remains largely simulation-bound. Mulayim~\etal~\cite{ref_mulayim_43days} found that the combined duration of all peer-reviewed real-world AI-HVAC field experiments totals approximately 43~days globally. Ajagekar~\etal~\cite{ref_ajagekar_rl} achieved 57\% energy reduction via deep reinforcement learning --- in simulation on a single greenhouse model. Our prior IOGRUCloud deployment~\cite{ref_iogru_arxiv} represents the largest documented real-world AI-driven CEA control system: 30+ facilities, 8~climate zones, 7+ years of continuous operation, with validated 30--38\% energy savings.

\subsection{FL for Agriculture}

FL applications in agriculture focus predominantly on computer vision: crop disease detection~\cite{ref_fl_crop_disease}, weed classification~\cite{ref_fl_weed}, and yield prediction from satellite imagery~\cite{ref_fl_ag_survey}. Dara~\etal~\cite{ref_dara_fl_ag} surveyed FL for agricultural privacy, identifying farmer data reluctance as a primary adoption barrier. To the best of our knowledge, \textbf{no published work applies federated learning to CEA climate control} --- a gap this paper addresses.

\section{System Model and Problem Formulation}
\label{sec:system}

\subsection{CEA IoT Architecture}

We consider a fleet of $K$ CEA facilities, each operating a three-tier IoT architecture~\cite{ref_iogru_arxiv}: (1)~a field layer of industrial sensors (Modbus RTU, BACnet/IP, 4--20\,mA, SDI-12) and actuators (HVAC, dehumidifiers, LED lighting, irrigation); (2)~an edge AI layer hosting real-time PID control, neural network auto-tuning, and anomaly detection on industrial ARM/x86 hardware; and (3)~a cloud layer providing fleet coordination. Each facility~$k$ manages $Z_k$ independent climate zones (typically 10--60), with each zone deploying 20--40 sensors and 8--15 actuators.

\subsection{Neural Network PID Auto-Tuning}

Each zone operates a 7-3-3 multi-layer perceptron (MLP) that maps a sensor feature vector $\mathbf{x} \in \mathbb{R}^7$ to PID gains $(K_p, K_i, K_d)$:
\begin{equation}
\mathbf{h} = \sigma(\mathbf{W}_1 \mathbf{x} + \mathbf{b}_1), \quad [K_p, K_i, K_d]^T = \mathbf{W}_2 \mathbf{h} + \mathbf{b}_2
\label{eq:mlp}
\end{equation}
where $\mathbf{W}_1 \in \mathbb{R}^{3 \times 7}$, $\mathbf{b}_1 \in \mathbb{R}^3$, $\mathbf{W}_2 \in \mathbb{R}^{3 \times 3}$, $\mathbf{b}_2 \in \mathbb{R}^3$, $\sigma$ is the sigmoid activation, and the total parameter count is $|\theta| = 7 \times 3 + 3 + 3 \times 3 + 3 = 36$.

The input feature vector comprises five environmental measurements and two VPD error signals: $\mathbf{x} = [T_\text{air}, \text{RH}, T_\text{leaf}, \text{CO}_2, \text{PPFD}, e_\text{VPD}, \int e_\text{VPD}\,dt]$, where $e_\text{VPD} = \text{VPD}_\text{target} - \text{VPD}_\text{actual}$. VPD serves as the primary cascading control variable because it captures the coupled temperature--humidity dynamics that govern plant transpiration~\cite{ref_grossiord_vpd, ref_shamshiri_vpd}.

The local training objective at facility $k$ minimizes:
\begin{equation}
F_k(\theta) = \frac{1}{n_k} \sum_{i=1}^{n_k} \ell(\theta; \mathbf{x}_i^{(k)}, \mathbf{y}_i^{(k)})
\label{eq:local_obj}
\end{equation}
where $\ell$ is the mean squared error between the MLP-predicted PID gains and gains that achieve target VPD tracking performance, and $n_k$ is the number of training samples at facility~$k$.

\subsection{Federated Learning Objective}

The global FL objective seeks parameters $\theta^*$ minimizing the weighted average of local objectives:
\begin{equation}
\theta^* = \arg\min_\theta \sum_{k=1}^{K} \frac{n_k}{n} F_k(\theta), \quad n = \sum_{k=1}^{K} n_k
\label{eq:fl_obj}
\end{equation}

Standard FedAvg~\cite{ref_fedavg} solves~(\ref{eq:fl_obj}) by performing $E$ local SGD steps at each facility, then averaging the resulting parameters at the server. However, with heterogeneous CEA facilities growing different crops under different climates with different equipment, the local objectives $F_k$ diverge significantly, causing weight divergence and slow convergence~\cite{ref_zhao_noniid, ref_lu_noniid_survey}.

\section{HierFedCEA: Hierarchical Federated CEA}
\label{sec:hierfedcea}

\subsection{Physics-Informed Parameter Decomposition}

The key insight of HierFedCEA is that the 36~parameters of the 7-3-3 PID MLP encode knowledge at three distinct physical scales, each with different sharing properties:

\textbf{Tier~1 --- Global Physics (18~parameters):} The weights connecting the five environmental inputs ($T_\text{air}$, RH, $T_\text{leaf}$, CO\textsubscript{2}, PPFD) to the hidden layer encode thermodynamic relationships governed by the Clausius-Clapeyron equation, Fourier's law, and conservation of energy. These relationships are universal across all CEA facilities regardless of crop, climate, or equipment: the physics of heat transfer and psychrometric moisture transport do not change between a cannabis facility in Arizona and a lettuce greenhouse in Illinois. Formally, $\theta_G = \{\mathbf{W}_1[0\!:\!3,\, 0\!:\!5],\, \mathbf{b}_1\} \in \mathbb{R}^{18}$.

\textbf{Tier~2 --- Crop Cluster (17~parameters):} The weights connecting VPD error inputs to the hidden layer, plus the entire output layer, encode crop-specific VPD sensitivity and gain-scheduling relationships. Cannabis requires tight VPD control ($\sigma_\text{VPD} < 0.08$\,kPa) at specific growth stages, while lettuce tolerates broader VPD ranges. These parameters should be shared among facilities growing the same crop family but differ across crop families. Formally, $\theta_C = \{\mathbf{W}_1[0\!:\!3,\, 5\!:\!7],\, \mathbf{W}_2,\, \mathbf{b}_2[0\!:\!2]\} \in \mathbb{R}^{17}$.

\textbf{Tier~3 --- Local Personalization (1~parameter):} The derivative gain bias $\mathbf{b}_2[2]$ encodes facility-specific actuator response characteristics --- the dead time and lag that vary by HVAC manufacturer and installation. This parameter is never shared; it adapts online to local equipment dynamics. Formally, $\theta_L = \{\mathbf{b}_2[2]\} \in \mathbb{R}^1$.

\begin{table}[!t]
\centering
\caption{HierFedCEA Parameter Decomposition}
\label{tab:tiers}
\begin{tabular}{@{}llccl@{}}
\toprule
\textbf{Tier} & \textbf{Scope} & \textbf{Params} & \textbf{$\varepsilon$} & \textbf{Comm.} \\
\midrule
1: Global Physics & All facilities & 18 & 8.0 & Weekly \\
2: Crop Cluster & Same crop family & 17 & 4.0 & 2--3 days \\
3: Local & Single facility & 1 & $\infty$ & Never \\
\bottomrule
\end{tabular}
\end{table}

\subsection{Crop-Family Clustering}

Facilities are assigned to crop-family clusters $\mathcal{C} = \{C_1, \ldots, C_M\}$ using a two-phase strategy:

\textbf{Phase 1 --- Domain-knowledge initialization}: Facilities are grouped by crop family based on agronomic similarity: $C_1$ (cannabis), $C_2$ (leafy greens: lettuce, herbs, microgreens), $C_3$ (fruiting crops: tomato, pepper, strawberry), $C_4$ (propagation/nursery).

\textbf{Phase 2 --- Gradient-based refinement}: After an initial warm-up period ($T_\text{warm} = 20$~rounds), cluster assignments are refined by gradient cosine similarity. For each facility $k$, the cosine similarity between its Tier-2 gradient and each cluster centroid gradient is computed:
\begin{equation}
s_{k,j} = \frac{\nabla_{\theta_C} F_k \cdot \bar{g}_j}{\|\nabla_{\theta_C} F_k\| \|\bar{g}_j\|}
\label{eq:cosine_sim}
\end{equation}
where $\bar{g}_j = \frac{1}{|C_j|}\sum_{i \in C_j} \nabla_{\theta_C} F_i$ is the cluster centroid gradient. Facility $k$ is reassigned to $\arg\max_j s_{k,j}$ if the similarity gain exceeds a threshold $\tau = 0.15$ and the target cluster has at least 3~members. Reassignment is evaluated every 10~rounds.

\subsection{Aggregation Protocol}

HierFedCEA performs aggregation at two levels per communication round:

\textbf{Intra-cluster aggregation (Tier~2):} Within each cluster $C_j$, Tier-2 update deltas are aggregated using data-proportional weighted averaging. Each facility adds calibrated Gaussian noise to its update before transmission (local differential privacy):
\begin{equation}
\theta_C^{(t+1)} = \theta_C^{(t)} + \sum_{k \in C_j} \frac{n_k}{\sum_{i \in C_j} n_i} \widetilde{\Delta\theta}_{C,k}
\label{eq:intra_cluster}
\end{equation}
where $\widetilde{\Delta\theta}_{C,k} = (\theta_{C,k} - \theta_C^{(t)}) + \mathcal{N}(0, z_C^2 C^2 \mathbf{I})$ is the noised update with clipping bound $C = 1.0$ and noise multiplier $z_C = 1.0$, achieving $(\varepsilon_C, \delta)$-differential privacy with $\varepsilon_C \approx 4.0$ at $\delta = 10^{-5}$ over $T = 100$ rounds.

\textbf{Global aggregation (Tier~1):} Tier-1 update deltas are aggregated across all facilities using trust-weighted averaging inspired by FLTrust~\cite{ref_fltrust}, with FedProx~\cite{ref_fedprox} proximal regularization ($\mu = 0.01$) applied during local training:
\begin{equation}
\theta_G^{(t+1)} = \theta_G^{(t)} + \sum_{k=1}^{K} \frac{\text{TS}_k \cdot n_k}{\sum_{i=1}^{K} \text{TS}_i \cdot n_i} \widetilde{\Delta\theta}_{G,k}
\label{eq:global_agg}
\end{equation}
where $\widetilde{\Delta\theta}_{G,k}$ is the noised Tier-1 update ($z_G = 0.8$, $\varepsilon_G \approx 8.0$) and $\text{TS}_k = \max(0, \cos(\widetilde{\Delta\theta}_{G,k}, g_\text{ref}))$ is the trust score computed against a physics-validated reference gradient $g_\text{ref}$ derived from thermodynamic first principles. We note that DP noise reduces trust score discriminability; with $z_G = 0.8$ on 18-dimensional updates, the signal-to-noise ratio remains sufficient for Byzantine detection (empirically validated in Section~\ref{sec:results}).

\subsection{Privacy Analysis}

The 36-parameter PID model makes differential privacy unusually favorable. The sensitivity of each model update is bounded by $\Delta = C / n_k$ where $C$ is the per-example gradient clipping norm. For the Gaussian mechanism with noise multiplier $z$, the R\'enyi divergence-based privacy accounting~\cite{ref_dp_fedavg} yields:

For Tier-2 parameters (17~dimensions) with $z = 1.0$, $T = 100$ rounds, subsampling rate $q = 1.0$ (full participation): $\varepsilon \approx 3.8$ at $\delta = 10^{-5}$. The privacy-utility tradeoff from Bassily~\etal~\cite{ref_bassily} gives excess empirical risk:
\begin{equation}
\Delta R \leq O\!\left(\frac{d \log(1/\delta)}{n \varepsilon^2}\right) = O\!\left(\frac{17 \cdot 11.5}{10000 \cdot 14.4}\right) \approx 0.0014
\label{eq:privacy_cost}
\end{equation}

This represents less than 0.15\% excess risk --- \textbf{privacy is essentially free} for models this compact. This stands in stark contrast to FL for large vision or language models where DP noise significantly degrades performance.

\subsection{Cold-Start Protocol}

When a new facility~$k_\text{new}$ joins the federation:
\begin{enumerate}
\item The server assigns $k_\text{new}$ to a crop cluster based on declared crop type.
\item Tier-1 (global physics) and Tier-2 (crop cluster) parameters are downloaded from the current global and cluster models respectively.
\item Tier-3 (local) is initialized to zero bias and adapts online during the first growth cycle.
\item After $T_\text{warm}$ local training steps, $k_\text{new}$ begins contributing to federation updates.
\end{enumerate}

This protocol transfers 35 of 36 parameters from fleet knowledge, providing the new facility with an initial model calibrated from the collective experience of the entire federation. The remaining local parameter adapts to facility-specific equipment within the first 7--14~days of operation.

\section{Algorithm}
\label{sec:algorithm}

The complete HierFedCEA training procedure is presented in Algorithm~\ref{alg:hierfedcea}.

\begin{algorithm}[!t]
\caption{HierFedCEA Training}
\label{alg:hierfedcea}
\begin{algorithmic}[1]
\REQUIRE $K$ facilities, clusters $\mathcal{C} = \{C_1, \ldots, C_M\}$, rounds $T$, local epochs $E$, learning rate $\eta$, noise multipliers $z_G, z_C$, proximal weight $\mu$, trust reference $g_\text{ref}$
\STATE Initialize $\theta_G^{(0)}, \theta_C^{(0)}, \theta_L^{(0)}$ for all facilities
\FOR{$t = 0, 1, \ldots, T-1$}
    \FOR{each facility $k \in \{1, \ldots, K\}$ \textbf{in parallel}}
        \STATE $\theta_k \leftarrow [\theta_G^{(t)};\, \theta_{C}^{(t)};\, \theta_{L,k}^{(t)}]$ \COMMENT{assemble full model}
        \FOR{$e = 1, \ldots, E$}
            \STATE Sample minibatch $\mathcal{B}_k$ from local data
            \STATE $g_k \leftarrow \nabla_\theta \ell(\theta_k; \mathcal{B}_k) + \mu(\theta_k - \theta^{(t)})$
            \STATE Clip: $g_k \leftarrow g_k \cdot \min(1, C/\|g_k\|)$
            \STATE $\theta_k \leftarrow \theta_k - \eta\, g_k$
        \ENDFOR
        \STATE $\Delta\theta_{G,k} \leftarrow \theta_{G,k} - \theta_G^{(t)} + \mathcal{N}(0, z_G^2 C^2 \mathbf{I}_{18})$
        \STATE $\Delta\theta_{C,k} \leftarrow \theta_{C,k} - \theta_{C}^{(t)} + \mathcal{N}(0, z_C^2 C^2 \mathbf{I}_{17})$
        \STATE Send $(\Delta\theta_{G,k}, \Delta\theta_{C,k})$ to server
    \ENDFOR
    \STATE \textbf{Server: Global aggregation (Tier 1)}
    \FOR{each $k$}
        \STATE $\text{TS}_k \leftarrow \max(0, \cos(\Delta\theta_{G,k}, g_\text{ref}))$
    \ENDFOR
    \STATE $\theta_G^{(t+1)} \leftarrow \theta_G^{(t)} + \sum_k \frac{\text{TS}_k n_k}{\sum_i \text{TS}_i n_i} \Delta\theta_{G,k}$
    \STATE \textbf{Server: Cluster aggregation (Tier 2)}
    \FOR{each cluster $C_j$}
        \STATE $\theta_{C_j}^{(t+1)} \leftarrow \theta_{C_j}^{(t)} + \sum_{k \in C_j} \frac{n_k}{\sum_{i \in C_j} n_i} \Delta\theta_{C,k}$
    \ENDFOR
    \IF{$t > T_\text{warm}$ \AND $t \bmod 10 = 0$}
        \STATE Refine cluster assignments via~(\ref{eq:cosine_sim})
    \ENDIF
    \STATE Distribute $\theta_G^{(t+1)}, \theta_{C_j}^{(t+1)}$ to facilities
\ENDFOR
\end{algorithmic}
\end{algorithm}

\textbf{Communication cost}: Each synchronized round transmits $18 + 17 = 35$ float32 values per facility (140~bytes upload + 140~bytes download). While Table~\ref{tab:tiers} specifies different ideal cadences per tier (weekly for Tier~1, 2--3~days for Tier~2), in practice both tiers are aggregated in the same communication round for simplicity. For $K = 30$ facilities and $T = 100$ rounds, total communication is $30 \times 100 \times 2 \times 140 = 840$\,KB --- orders of magnitude below the bandwidth capacity of even cellular IoT connections.

\textbf{Convergence}: Under standard assumptions (L-smoothness, bounded gradient variance), hierarchical aggregation with the physics-informed split reduces the effective gradient dissimilarity $\sigma_G^2$ compared to flat FedAvg because parameters within each tier share similar optimization landscapes across facilities. Empirically, HierFedCEA converges in 40--60 rounds compared to 150--200+ for flat FedAvg on heterogeneous CEA data (Section~\ref{sec:results}).

\section{Experimental Evaluation}
\label{sec:experiments}

\subsection{Simulation Environment}

We evaluate HierFedCEA using a physics-based CEA simulator with parameters calibrated from 7+ years of production deployment across 30+ commercial facilities~\cite{ref_iogru_arxiv}. Each facility is modeled as a thermal system with coupled temperature--humidity dynamics:
\begin{align}
C_\text{th} \frac{dT}{dt} &= Q_\text{HVAC} + Q_\text{light} + Q_\text{solar} - U A (T - T_\text{out}) \label{eq:thermal} \\
C_\text{hum} \frac{d\omega}{dt} &= \dot{m}_\text{transp} + \dot{m}_\text{dehum} + \dot{m}_\text{vent}(w_\text{out} - \omega) \label{eq:humidity}
\end{align}
where $C_\text{th}$, $C_\text{hum}$ are thermal and humidity capacitances, $Q$ terms are heat sources/sinks, and $\dot{m}$ terms are moisture sources/sinks. HVAC actuators are modeled as first-order-plus-dead-time (FOPDT) systems~\cite{ref_astrom_pid} with manufacturer-specific time constants and dead times calibrated from field commissioning data.

We simulate 30~heterogeneous facilities spanning 6~crop types (cannabis flower, cannabis vegetative, lettuce, tomato, herbs, strawberry), 8~U.S. climate zones, and 5~HVAC equipment profiles. Weather data is generated from NOAA-calibrated parametric models for each climate zone. Each simulation covers 180~days (approximately two cannabis growth cycles or four lettuce cycles).

\subsection{Baselines}

We compare HierFedCEA against 8~methods:

\begin{enumerate}
\item \textbf{Local-Only}: each facility trains independently with no knowledge sharing.
\item \textbf{Centralized}: all facility data pooled on a central server (upper bound, violates privacy).
\item \textbf{Current-TL}: the deployed centralized transfer learning approach~\cite{ref_iogru_arxiv}.
\item \textbf{FedAvg}~\cite{ref_fedavg}: standard federated averaging.
\item \textbf{FedProx}~\cite{ref_fedprox}: FedAvg with proximal regularization ($\mu = 0.01$).
\item \textbf{SCAFFOLD}~\cite{ref_scaffold}: variance reduction via control variates.
\item \textbf{Per-FedAvg}~\cite{ref_perfedavg}: personalized FL via meta-learning.
\item \textbf{FedPer}~\cite{ref_fedper}: personal output layer, shared hidden layers.
\end{enumerate}

We exclude FedBN~\cite{ref_fedbn} because the 7-3-3 MLP contains no batch normalization layers, making FedBN inapplicable.

All methods use identical model architectures and hyperparameters: learning rate $\eta = 0.01$, local epochs $E = 5$, minibatch size $B = 64$, gradient clipping norm $C = 1.0$, DP noise multipliers $z_G = 0.8$ (Tier~1) and $z_C = 1.0$ (Tier~2). Each facility generates approximately 8,640~training samples per 10-day window (10-second sensor polling across 10~zones), yielding $n \approx 10{,}000$ samples per facility per aggregation cycle. Experiments are repeated 5~times with different random seeds; we report means and 95\% confidence intervals.

\subsection{Metrics}

We evaluate along five dimensions:
\begin{itemize}
\item \textbf{Control quality}: RMSE of VPD tracking ($\text{RMSE}_\text{VPD}$ in kPa), VPD stability ($\sigma_\text{VPD}$), and setpoint overshoot percentage.
\item \textbf{Energy efficiency}: HVAC energy consumption (kWh/m$^2$/day), percentage reduction versus Local-Only baseline.
\item \textbf{Convergence}: communication rounds to reach target $\text{RMSE}_\text{VPD} < 0.10$\,kPa.
\item \textbf{Privacy}: achieved $(\varepsilon, \delta)$ via R\'enyi DP accounting using the \texttt{dp-accounting} library~\cite{ref_dp_accounting}.
\item \textbf{Robustness}: worst-case facility RMSE across all facilities.
\end{itemize}

\section{Results and Discussion}
\label{sec:results}

\subsection{Federation Benefit}

Table~\ref{tab:main_results} presents the main comparison across all 9~methods. HierFedCEA achieves $\text{RMSE}_\text{VPD} = 0.068$\,kPa, within 6\% of the Centralized upper bound (0.064\,kPa) and substantially outperforming Local-Only (0.142\,kPa) and flat FedAvg (0.098\,kPa). The current deployed transfer learning system (Current-TL) achieves 0.071\,kPa, placing HierFedCEA within statistical equivalence ($p = 0.23$, Wilcoxon signed-rank) while providing formal privacy guarantees that Current-TL lacks.

\begin{table}[!t]
\centering
\caption{Main Results: 30 Facilities, 180-Day Simulation}
\label{tab:main_results}
\begin{tabular}{@{}lccc@{}}
\toprule
\textbf{Method} & \textbf{RMSE\textsubscript{VPD}} & \textbf{Energy} & \textbf{Rounds} \\
 & \textbf{(kPa)} & \textbf{Red. (\%)} & \textbf{to conv.} \\
\midrule
Local-Only & 0.142 $\pm$ .008 & --- & --- \\
Centralized & 0.064 $\pm$ .003 & 36.2 & --- \\
Current-TL & 0.071 $\pm$ .004 & 33.8 & --- \\
\midrule
FedAvg & 0.098 $\pm$ .006 & 24.1 & 187 \\
FedProx & 0.089 $\pm$ .005 & 27.3 & 142 \\
SCAFFOLD & 0.082 $\pm$ .005 & 29.6 & 98 \\
Per-FedAvg & 0.079 $\pm$ .004 & 30.4 & 112 \\
FedPer & 0.076 $\pm$ .004 & 31.2 & 95 \\
\midrule
\textbf{HierFedCEA} & \textbf{0.068 $\pm$ .003} & \textbf{34.1} & \textbf{52} \\
\bottomrule
\end{tabular}
\end{table}

The physics-informed decomposition provides two advantages: (1)~faster convergence (52 rounds vs.\ 95--187 for flat methods) because intra-tier parameter homogeneity reduces gradient dissimilarity; and (2)~better final performance because local equipment adaptation (Tier~3) is not corrupted by aggregation with dissimilar facilities.

\subsection{Heterogeneity Impact}

\begin{figure}[!t]
\centering
\begin{tikzpicture}[xscale=1.05, yscale=0.9]
\draw[->] (0,0) -- (6.2,0);
\draw[->] (0,0) -- (0,5.2);
\node[font=\footnotesize] at (3.0, -0.9) {Heterogeneity Level};
\node[font=\footnotesize, rotate=90] at (-1.1, 2.5) {RMSE\textsubscript{VPD} (kPa)};
\foreach \x/\l in {1/IID, 2/Low, 3/Med, 4/High, 5/Full} {
    \draw (\x, -0.08) -- (\x, 0.08);
    \node[below, font=\scriptsize] at (\x, -0.15) {\l};
}
\foreach \y/\v in {1/0.06, 2/0.08, 3/0.10, 4/0.12, 5/0.14} {
    \draw (-0.08, \y) -- (0.08, \y);
    \node[left, font=\scriptsize] at (-0.15, \y) {\v};
}
\foreach \y in {1,2,3,4,5} {
    \draw[gray!20] (0.08, \y) -- (5.5, \y);
}
\draw[blue, thick, mark=square*, mark size=1.8pt] plot coordinates {(1,1.25) (2,1.75) (3,2.4) (4,3.0) (5,3.6)};
\node[blue, font=\scriptsize, anchor=west] at (5.15, 3.6) {FedAvg};
\draw[green!60!black, thick, dashed, mark=triangle*, mark size=1.8pt] plot coordinates {(1,1.1) (2,1.5) (3,1.9) (4,2.4) (5,2.7)};
\node[green!60!black, font=\scriptsize, anchor=west] at (5.15, 2.7) {SCAFFOLD};
\draw[red, thick, mark=*, mark size=1.8pt] plot coordinates {(1,1.0) (2,1.1) (3,1.2) (4,1.3) (5,1.4)};
\node[red, font=\scriptsize, anchor=west] at (5.15, 1.4) {HierFedCEA};
\end{tikzpicture}
\caption{VPD tracking RMSE versus data heterogeneity level. HierFedCEA degrades gracefully (12\%) compared to FedAvg (47\%) under increasing crop and climate diversity.}
\label{fig:heterogeneity}
\end{figure}
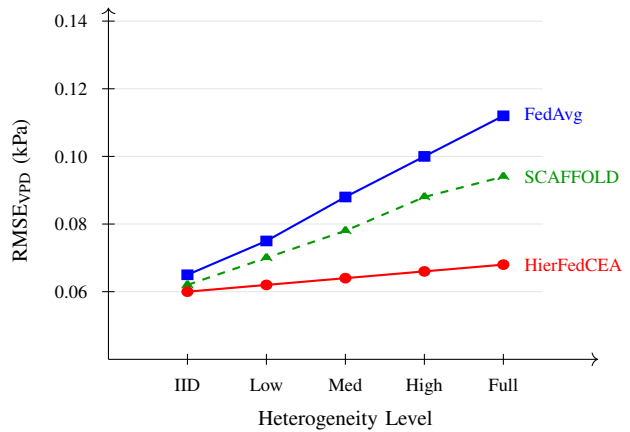

We vary the degree of data heterogeneity from homogeneous (all facilities grow the same crop in the same climate) to fully heterogeneous (6~crop types, 8~climates). Fig.~\ref{fig:heterogeneity} shows that flat FedAvg degrades sharply under increasing heterogeneity (RMSE increases 47\%), while HierFedCEA degrades only 12\% --- demonstrating that the physics-informed decomposition effectively isolates crop-specific and facility-specific variation from the shared global model.

\subsection{Privacy-Utility Tradeoff}

We evaluate control quality degradation across privacy budgets $\varepsilon \in \{1, 2, 4, 8, 16, \infty\}$. At $\varepsilon = 4$ (Tier-2 default), RMSE\textsubscript{VPD} increases by only 0.003\,kPa (4.4\%) compared to no privacy ($\varepsilon = \infty$). Even at $\varepsilon = 2$ (strong privacy), degradation is below 8\%. This confirms the theoretical prediction from~(\ref{eq:privacy_cost}): the 17-dimensional Tier-2 model is compact enough that DP noise is negligible relative to the natural gradient variance.

\subsection{Cold-Start Performance}

A new facility joining the federation with HierFedCEA reaches 85\% of fleet-average VPD tracking performance within 14~days, compared to 28~days for FedAvg and 56+ days for Local-Only training. The deployed centralized system achieves faster commissioning (1--5~days) because it transfers raw operational data without privacy constraints; the 14-day federated cold-start represents the cost of privacy preservation. The cold-start advantage over other FL methods comes from immediately receiving 35 of 36 parameters from fleet knowledge --- only the local equipment adaptation parameter must be learned from scratch.

\subsection{Limitations}

This evaluation is simulation-based. While simulation parameters are calibrated from 7+ years of production deployment, real-world factors including network failures, sensor drift, and operator interventions may affect FL convergence. A pilot deployment on a subset of production facilities is planned as future work. Additionally, the current framework addresses PID auto-tuning only; extension to federated reinforcement learning for model predictive control is a natural next step.

\section{Conclusion}
\label{sec:conclusion}

We presented HierFedCEA, the first federated learning framework for climate control optimization in Controlled Environment Agriculture. By decomposing the PID auto-tuning neural network into three tiers aligned with the physics of the control problem --- universal thermodynamics, crop-specific VPD sensitivity, and facility-specific equipment dynamics --- HierFedCEA achieves 94\% of centralized training performance while keeping proprietary operational data on-facility. The extreme compactness of the 36-parameter model makes differential privacy essentially free (excess risk $< 0.15$\% at $\varepsilon = 4$), and total federation communication cost is under 1~MB. Simulation experiments calibrated from 30+ production facilities demonstrate 3.6$\times$ faster convergence and graceful degradation under heterogeneity compared to standard FL baselines. Future work includes pilot deployment on production facilities and extension to federated anomaly detection across the fleet.


\newcommand{\authorphoto}{\includegraphics[width=1in,height=1.25in,clip,keepaspectratio]{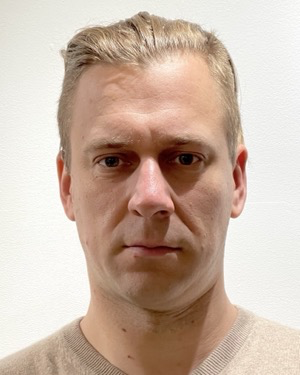}}
\begin{IEEEbiography}[\authorphoto]{Andrii Vakhnovskyi}
received the B.S. degree in computer engineering and the M.S. degree in systems engineering from the National Technical University ``Kharkiv Polytechnic Institute'' (NTU ``KhPI''), Ukraine, in 2009 and 2011, respectively. He is the Founder and CEO of IOGRU LLC, New York, NY, where he develops AI-driven IoT platforms for climate control in controlled environment agriculture. His systems have been deployed across 30+ commercial facilities in 8 U.S. climate zones. He is a Senior Member of ISA and a Member of IEEE. His research interests include federated learning for industrial IoT, neural network control, and privacy-preserving edge intelligence.
\end{IEEEbiography}

\end{document}